# Optical intraday variability analysis for the BL Lacertae object 1ES 1426+42.8

X. Chang ,[1] D. R. Xiong ,[2]★ T. F. Yi ,[3,4]★ C. X. Liu ,[1] G. Bhatta ,[5] J. R. Xu ,[6]★ and Y. L. Gong [7]

[1]*South-Western Institute for Astronomy Research, Yunnan University, Kunming 650500, People's Republic of China*  
[2]*Yunnan Observatories, Chinese Academy of Sciences, 396 Yangfangwang, Guandu District, Kunming 650216, People's Republic of China*  
[3]*Key Laboratory of Colleges and Universities in Yunnan Province for High-energy Astrophysics, Department of Physics, Yunnan Normal University, Kunming 650500, People's Republic of China*  
[4]*Guangxi Key Laboratory for the Relativistic Astrophysics, Nanning 530004, People's Republic of China*  
[5]*Janusz Gil Institute of Astronomy, University of Zielona Góra, ul. Szafrana 2, PL-65-516 Zielona Góra, Poland*  
[6]*Shandong Provincial Key Laboratory of Optical Astronomy and Solar-Terrestrial Environment, School of Space Science and Physics, Institute of Space Sciences, Shandong University, Weihai 264209, People's Republic of China*  
[7]*Department of Astronomy, School of Physics and Astronomy, Yunnan University, Kunming 650091, People's Republic of China*

## ABSTRACT

The observation data of blazar 1ES 1426+42.8 were obtained using the 1.02 m optical telescope of Yunnan Observatories during 2021 to 2023. Intraday variability (IDV) is detected on seven nights. We use the turbulent model to investigate the mechanism of IDV in 1ES 1426+42.8. The fitting light curves match the actual IDV curves well. Using this model, we obtain the parameters such as the size of turbulent cells and the width of pulses in the jet. A possible short-lived quasi-periodic oscillation (QPO) of $58.55 \pm 8.09$ min was detected on 2022 April 26 whose light curve exhibits eight cycles at $> 3\sigma$ global significance and confirmed by several different techniques. Through a more detailed analysis of the light curve of this night, we find that the period is shortened from 54.23 min ($4\sigma$) to 29.71 min ($3\sigma$). The possible QPO and period shortening phenomenon are best explained by the processes of magnetic reconnections.

## 1 INTRODUCTION

Blazars are the most energetic populations of active galactic nuclei (AGNs) with relativistic jets pointing at a small angle to the line of sight. They are characterized by rapid flux variations throughout the entire electromagnetic spectrum, high luminosities, and strong polarizations (Urry & Padovani 1995). Studies of blazars can help us understand the mechanisms that power the large-scale jets of AGNs as well as the physical properties of accretion discs near central supermassive black holes (SMBHs). Blazars consist of the BL Lacertae objects (BL Lacs) with almost featureless continuum spectra, and the flat spectrum radio quasars with prominent emission lines (Agarwal & Gupta 2015).

The study of variability is one of the most powerful tools for understanding the processes occurring in blazars. The rapid and significant variability observed in blazars can be utilized to investigate the internal physical processes of jets, including particle acceleration, the origin of flares, and the structure/dynamics of emission regions (Miller, Carini & Goodrich 1989; Gupta, Srivastava & Wiita 2009;

Liu & Bai 2015; Xiong et al. 2017). Extremely rapid variation over a few minutes to less than a day is often called intraday variability (IDV) or microvariability (Miller et al. 1989; Agarwal & Gupta 2015). Some blazars exhibit high amplitudes of variability across different wavebands, occurring on time-scales as short as several minutes (Sagar et al. 2004). Models related to jets and accretion discs have been proposed to explain the variability behaviors observed in blazars, but numerous details of these models are still under discussion (Bhatta 2021). Webb et al. (2021) present a theory to explain the IDV/microvariability. This model assumes that the observed flares are caused by shock waves propagating down the turbulent jet. Particles in each turbulence plasma cell are accelerated by the shocks and subsequently cooled by synchrotron radiation, resulting in pulses or flares in the light curves (Błażejowski et al. 2000; Xu et al. 2019; Webb et al. 2021). The turbulent cells have specific sizes, densities, and magnetic field orientations. The flare duration indicates the size of the turbulent cells, and the amplitude correlates with the density and magnetic field properties of the cells. Solutions to the particle kinetic equations with synchrotron cooling can predict the shape of pulses emitted at specific frequencies as a result of the shock wave interacting with the turbulent cells (Kirk, Rieger & Mastichiadis 1998; Webb et al. 2017; Webb & Sanz 2023). By deconvolving the light curves into individual pulses and analysing the flare distribution characteristics, the turbulence

★ E-mail: xiongdingrong@ynao.ac.cn (DX); yitingfeng98@163.com (TY); xujingrichard@foxmail.com (JX)

characteristics in the jet can be studied (Bhatta et al. 2013; Webb et al. 2021).

In blazars, jets originate from the central engines of SMBHs. The IDVs are most likely produced in the vicinity of the SMBHs (e.g. Miller et al. 1989). Quasi-periodic oscillations (QPOs) on intraday time-scales are rarely detected in blazars (Hong, Xiong & Bai 2018). In the past decade, researchers have used data from different wavebands to detect QPOs in blazars across various time-scales. Additionally, the significance of the signals of QPOs still need to be further confirmed (Gupta et al. 2009; Lachowicz et al. 2009; Rani et al. 2010). Gupta et al. (2009) detected IDV QPOs with periods of ∼ 25–73 min for S5 0716+714 with high probabilities (95 per cent– $\gtrsim$ 99 per cent). Hong et al. (2018) confirmed their period estimation of the S5 0716+714 QPO to be ∼50 min at the 99 per cent significance level using the z-transformed discrete correlation function method and the Lomb–Scargle method. Rani et al. (2010) also discovered a QPO with a period of ∼15 min at the 3$\sigma$ confidence level in S5 0716+714 using several different techniques. The potential explanation for the observed IDV QPO is magnetic reconnection in the jet systems. In the inner regions of the jet flows, magnetic reconnection can be generated by magnetohydrodynamic turbulence, causing periodic instabilities and quasi-periodic fluctuations in the relativistic jets (Balbus & Hawley 1991; Čemeljić et al. 2022). Jet systems are dominated by kink instabilities, which can more efficiently generate magnetic reconnection, and result in quasi-periodic releases of emission energy (Dong, Zhang & Giannios 2020; Jorstad et al. 2022; Raiteri et al. 2023). The observed periods can be drastically shortened by the relativistic effects.

The blazar 1ES 1426+42.8 was discovered in the medium X-ray band (2–6 keV) with the Large Area Sky Survey experiment (LASS) on High Energy Astronomical Observatory (HEAO-1) (Wood et al. 1984). Abdo et al. (2010) classified 1ES 1426+42.8 ($z = 0.129$) as a BL Lac object based on its featureless spectral energy distributions. In addition, it is an extreme TeV source with low power and high synchrotron peak ($\nu_{peak} > 100$ keV) (Costamante et al. 2003). Presently, most of the observations and explorations of 1ES 1426+42.8 focus on the high-energy bands, while the optical bands remain less studied. Gaur et al. (2010) collected a sample of blazars with IDV light curves that have at least four peaks to search for potential IDV QPOs. They found a possible QPO in one light curve of PKS 2155−304, but no QPO was found in 1ES 1426+42.8. However, Chang et al. (2023) detected a potential QPO with a period of 48.67 ± 13.90 min in 1ES 1426+42.8 on 2010 April 13, and this QPO is further confirmed at > 3$\sigma$ level on 2021 March 16. Unfortunately, the number of data points available for the QPO light curves in these 2 d is limited. With undersampled light curves, the confidence level is relatively low. The motivation of this work is to search for IDV and further investigate the existence of QPO in 1ES 1426+42.8 with higher confidence levels, and to explore the physical mechanisms of the IDV and QPO. The study of the IDV and QPO can provide deeper insights of accretion discs and jets, as well as the dynamic processes that occur in the central engines of blazars (Li et al. 2023). Therefore, it would be beneficial to search for IDV and QPO in the light curves of blazars as many as possible. In this work, we conducted new optical monitoring for 1ES 1426+42.8 from 2021 to 2023 (16 nights), and present the results of our IDV and QPO study.

This paper is structured as follows. Section 2 describes the observations and data analyses. Section 3 describes the results of variability and periodicity analyses. Our discussions and conclusions are presented in Sections 4 and 5.

## 2 OBSERVATIONS AND DATA REDUCTION

Our observations of 1ES 1426+42.8 were carried out with the 1.02 m optical telescope administered by Yunnan Astronomical Observatories of China. During our observations, the 1.02 m telescope was equipped with an Andor Ikon XL CCD (4096 × 4112 pixels) camera at the Cassegrain focus ($f = 13.3\,m$). The field of view is 15 × 15 arcmin$^2$. The pixel scale is 0.238 arcsec pixel$^{-1}$. The standard Johnson broad-band filters are used. Our photometric observations were performed through the *I*-band filter. A 2-min exposure time was adopted to achieve a high signal-to-noise ratio (SNR) and ensure intensive sampling. Five sky flat frames were taken at dusk, and 10 bias frames were captured immediately before the scientific observations. The APPHOT task of the IRAF[1] software is used for aperture photometric measurements with flats and biases corrected. Aperture magnitudes are calculated through a set of radii. The aperture with a radius of twice of the full width at half-maximum (FWHM) of the point spread function, $r = 2 \times FWHM$, is finally adopted for the best SNR. We measured the instrumental magnitudes of the comparison star and the check star in the same field of 1ES 1426+42.8. The comparison star and the check star are taken from the finding chart of Smith, Jannuzi & Elston (1991)[2] for flux calibration. The comparison star is required to have the colour and the brightness similar to 1ES 1426+42.8. The check star is chosen to be the one with the smallest variations in differential magnitudes with respect to the comparison star (Agarwal & Gupta 2015; Xiong et al. 2017). Then we chose to use Star 1 as the comparison star and measure Star 2 as the check star (see the finding chart of Smith et al. 1991). The apparent magnitude of 1ES 1426+42.8 can then be calculated from the comparison star and the check star using differential photometry (Bai et al. 1998; Fan et al. 2014). According to the magnitudes of the comparison star and the check star, the root-mean-square (RMS) errors of the photometry at this night can be calculated (Xiong et al. 2017):

$$\sigma = \sqrt{\frac{\sum(m_i - \overline{m})^2}{N - 1}}, \tag{1}$$

where $m_i$ represents the differential magnitude of the comparison star and check star, $\overline{m}$ represents the averaged differential magnitude for one night, and $N$ represents the total number of observations on a given night.

We conducted the monitoring program for 1ES 1426+42.8 from 2021 to 2023 in I band with additional continuous observations on 2021 April 22 in C band (clear band). Excluding the nights with bad weather, the total number of nights with continuous observations for 1ES 1426+42.8 is 16 (1818 data points). They are January 25, March 16, 23, and 24, April 22 and 29, May 18 in 2021, April 24–27 in 2022, and April 17–21 in 2023. Details of the observations are listed in Table 1. The light curves from 2021 to 2023 are presented in Figs 1–3, respectively.

---

[1] IRAF is distributed by the National Optical Astronomy Observatories, which are operated by the Association of Universities for Research in Astronomy, Inc., under cooperative agreement with the National Science Foundation.
[2] https://www.lsw.uni-heidelberg.de/projects/extragalactic/charts/1426+428.html

**Table 1.** The log of observations. Columns 1 to 5 represent the Universal Time (UT), Julian Day (JD), magnitudes, RMS errors, and the observed band, respectively.

| Date (UT) | JD | Mag | $\sigma$ | Band |
|---|---|---|---|---|
| 2021 Jan 25 | 2459240.398 | 15.626 | 0.047 | I |
| 2021 Jan 25 | 2459240.400 | 15.653 | 0.047 | I |
| – | – | – | – | – |

*Note.* (This table is available in its entirety in machine-readable form.)

## 3 RESULTS

### 3.1 Variability analysis

In order to quantify the IDV/microvariation, we employed the Howell statistical method (Howell, Mitchell & Warnock 1988) (hereafter Howell Test). The Howell Test compares the variances of the object to the variances of a comparison star and of a check star, accounting for the detector's instrumental noise properties and brightness differences between the source and comparison star, to determine whether observed variations are real or merely statistical noise (Howell et al. 1988; de Diego 2010; Webb et al. 2021). Based on the characteristics of CCD, it computes the probability that the detected variation in the variable is real, rather than random noise (Howell et al. 1988). The $F$ value in the Howell Test is calculated as follows:

$$F = \frac{S^2_{(BL-StarA)}}{\Gamma^2 S^2_{(StarA-StarB)}}, \quad (2)$$

where $S^2_{(BL-StarA)}$ is the measured variance of the differential instrumental magnitudes between the blazar $BL$ and comparison $StarA$, while $S^2_{(StarA-StarB)}$ is the measured variance of the differential instrumental magnitudes between comparison $StarA$ and check $StarB$. $\Gamma^2$ is a statistical correction factor calculated based on the known properties of CCD, which is used to correct the different SNRs for the objects in CCD frame due to the photon noise (Howell et al. 1988). The $F$ value is compared with the critical $F$ value, $F^\alpha_{\nu_{bl},\nu_*}$, where $\nu_{bl}$ and $\nu_*$ represent the number of degrees of freedom for the blazar and comparison star, respectively ($\nu = N-1$), and $\alpha$ represents the significance level set as 0.01 ($2.6\sigma$) (Xiong et al. 2017). If the $F$ value exceeds the critical value, the blazar can be considered variable with a 99 per cent confidence level. Another alternative to the standard F-test is the one-way analysis of variance (ANOVA) test (de Diego 2010). de Diego (2010) reported that ANOVA is a powerful and robust estimator for microvariations. The ANOVA method derives the expected variance from subsamples of the data, rather than relying on error measurement. Considering the exposure time, we bin the data in groups of three or five observations (de Diego 2010; Xiong et al. 2017). If measurements in the last group are less than three or five, then they will be combined with the previous group. $F^\alpha_{\nu_1,\nu_2}$ in the F-statistics can be used to obtain the critical value of ANOVA, where $\nu_1 = k-1$ ($k$ is the number of groups), $\nu_2 = N-k$ ($N$ is the number of measurements), and $\alpha$ is the significance level (Hu et al. 2014). A blazar is classified as variable (V) if its light curve passes both tests on a given night. If one or all of the criteria are not met, the blazar is classified as non-variable (N). Table 2 displays the results of the variability analysis. The results show that IDVs were found on seven nights (2021 March 16 and 24, 2022 April 24–26, 2023 April 17 and 21). The IDVs in this object are intermittent, and there is no correlation between brightness level and the presence or absence of IDV.

### 3.2 Periodicity analysis

Blazars' QPO research holds significant values in terms of investigating the physical mechanisms as well as radiation processes in blazars (Gupta et al. 2009). Figs 1, 2, and 3 present the 16 light curves observed in the year of 2021, 2022, and 2023, respectively. The light curve on 2021 March 16 shows four obvious peaks (the upper right panel of Fig. 1). There are eight cycles in the light curve of 2022 April 26 (the bottom left panel of Figs 2 and 5). This light curve is considered to exhibit IDV (with an amplitude of 61.08 per cent) because it follows the normal distribution and it successfully passes both the F-test and the one-way ANOVA test (de Diego 2010; Joshi et al. 2011). Thus, we conduct a detailed QPO analysis on the light curve of 2022 April 26.

We first use Weighted Wavelet Z-transform (WWZ: see Section 3.2.1) method to search for QPO for the whole optical light curve. The analysis result is presented in Fig. 4. Beyond the initial stable variations before JD (2459600+) 96.144, there is a bright red patch from JD(2459600+) 96.144 to 96.365 (segment 1) in Fig. 4, which represents an obvious QPO signal. For the segment 1, signal of brightest red patch appears between JD(2459600+) 96.144 and 96.286 (segment 2). Thus, we select these two segments to explore QPOs with three methods: the Lomb–Scargle method, WWZ, and the REDFIT method.

#### 3.2.1 Lomb–Scargle periodogram and WWZ

Lomb–Scargle periodogram (LSP) (Lomb 1976; Scargle 1982) is a method widely used to search for and analyse QPOs. LSP competes the traditional Discrete Fourier Transform by employing a least-squares fitting technique to approximate sinusoidal waves in the data to reduce the effect of uneven sampling (Lomb 1976; Scargle 1982; Bhatta 2017). The periodogram is a function of circular frequency $f$ (Li et al. 2016). The power spectrum density function of a real signal exhibits a peak, and the maximum power at that peak frequency corresponds to the most probable period (Wang et al. 2014). The WWZ (Foster 1996; Bhatta 2017) method is utilized for detecting and quantifying QPOs in both the frequency and time domains. The frequencies and amplitudes of QPOs may vary over time in real astronomical systems. In such cases, the WWZ method is particularly effective in the identifications of QPOs that develop and dissipate with time (Bhatta 2017). A set of Morlet wavelets with different combinations of positions $\tau$ (time shifts) and scale frequencies $\omega$ (test frequencies) are used to analyse and construct the WWZ spectra (Foster 1996; Wang et al. 2014; Bhatta 2017). The significance of a detected periodicity is estimated by the WWZ power in a statistical manner (Foster 1996). The WWZ power peaks can be used to determine periodic components and track the evolution of periodicities and/or amplitudes (Li et al. 2021).

The transitory apparently periodic fluctuations of blazars usually have higher level of red noises at lower frequencies or flicker noises, which can result in spurious periodicities generated in the periodograms (Press 1978). When analysing the periodicity of blazars, it is necessary to carefully consider the effects of frequency-dependent red-noise or flicker-noise (Vaughan 2005; Fan et al. 2014; Sandrinelli et al. 2016; Bhatta 2017). This issue can be solved using the power response method, which characterizes the power spectral density (PSD) (Uttley, McHardy & Papadakis 2002). The random fluctuations of blazars are generally modelled as a power-law PSD with the form $P(f) \propto f^{-\alpha}$, where $P(f)$ represents the power at temporal frequency $f$ and $\alpha$ is the spectral slope. Following Vaughan (2005), we used the linear regression to estimate the power spectral slope $\alpha$ by fitting a linear function to the log-periodogram.

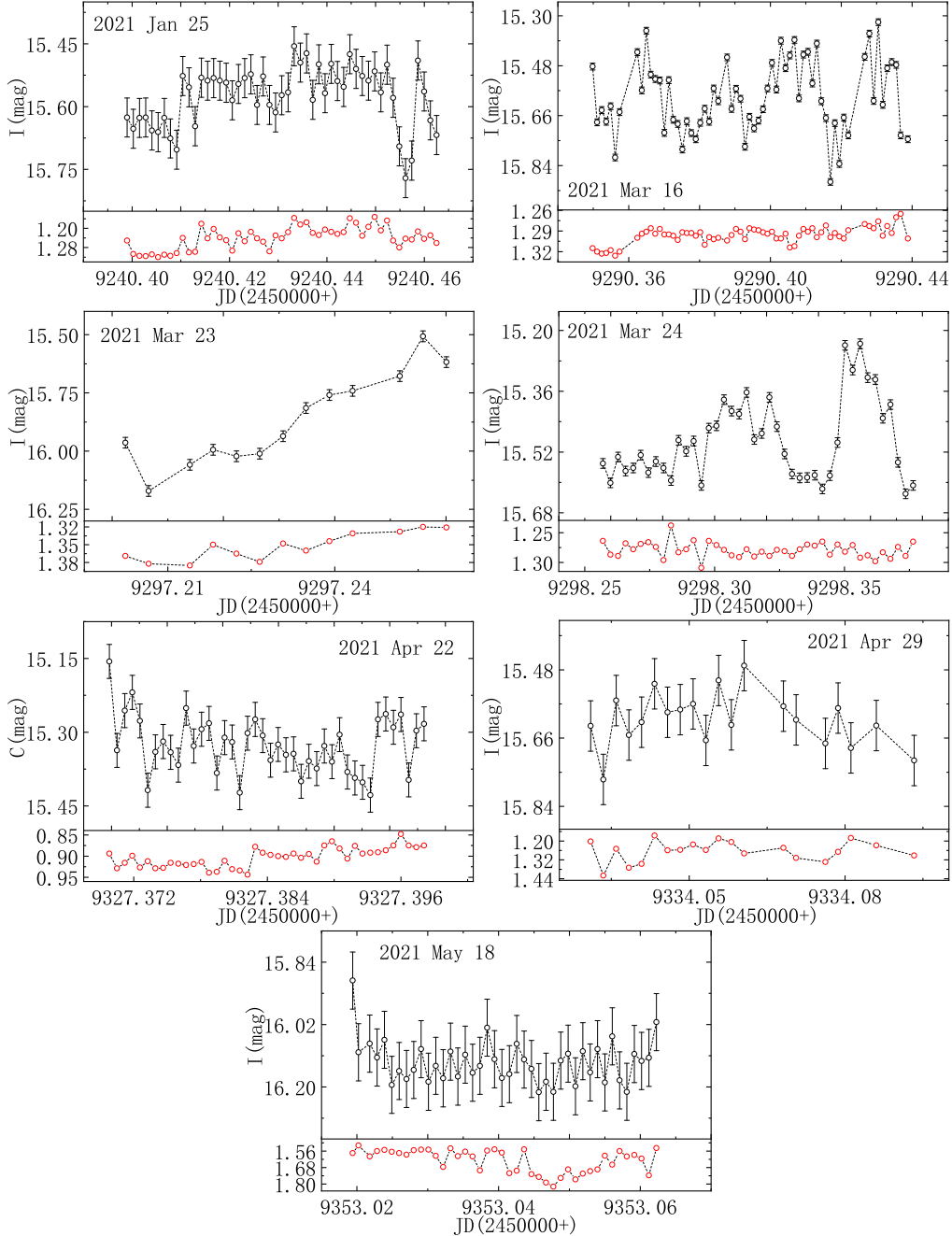

**Figure 1.** Light curves of 1ES 1426+42.8 in 2021. The black open circles are the light curves for the sources. The red open circles are the magnitude difference between comparison stars and check stars in the same period.

The left panels of Figs 6 and 7 display the best-fitting PSDs of segment 1 and segment 2 obtained from this process. The PSD analysis gives the slopes of $\alpha = 1.21 \pm 0.02$ and $\alpha = 1.02 \pm 0.03$. This value is then used to model the red noises in the optical variability of 1ES 1426+42.8. The next step is to assess the confidence level of the QPO. We generate 100 000 simulated light curves based on the best-fitting PSD model and then used the Monte Carlo method to establish the red-noise background as described in Timmer & Koenig (1995). These artificial light curves have identical sampling interval, standard deviation, and mean value. The LSP and WWZ power spectrum can be obtained for each simulated light curve (Li et al. 2023). The confidence levels of 99.7 per cent, 99.99 per cent, and 99.994 per cent ($4\sigma$) can then be estimated using the power spectral distribution of the simulated light curves. The results of LSP and WWZ are shown in middle and right panels of Figs 6 and 7, respectively. The red, blue, and purple lines in Fig. 6 represent confidences of 99 per cent, 99.7 per cent, and 99.9 per cent, respectively, while the lines in Fig. 7 represent 99.7 per cent, 99.9 per cent, and 99.994 per cent confidence, respectively. We use the FWHM of the peak as the uncertainty of the periodic modulation. As shown in middle panel of Fig. 6, there is an obvious peak (marked by the red arrow) in the periodogram of LSP for segment 1. This suggests a potential QPO with a period of $58.55 \pm 8.09$ min at > 99.9 per cent for 1ES 1426+42.8 on 2022 April 26. Similarly, the highest WWZ power peak in the right panel of Fig. 6 indicates a period of $58.34 \pm 7.78$ min at > 99.7 per cent ($3\sigma$). Fig. 7 suggests a similar QPO with higher confidence in segment 2 with a period of $54.23 \pm 10.73$ min (> $4\sigma$) using the LSP method, and a period of $53.88 \pm 9.71$ min at > 99.994 per cent ($4\sigma$) using the WWZ method. Both methods give consistent results.

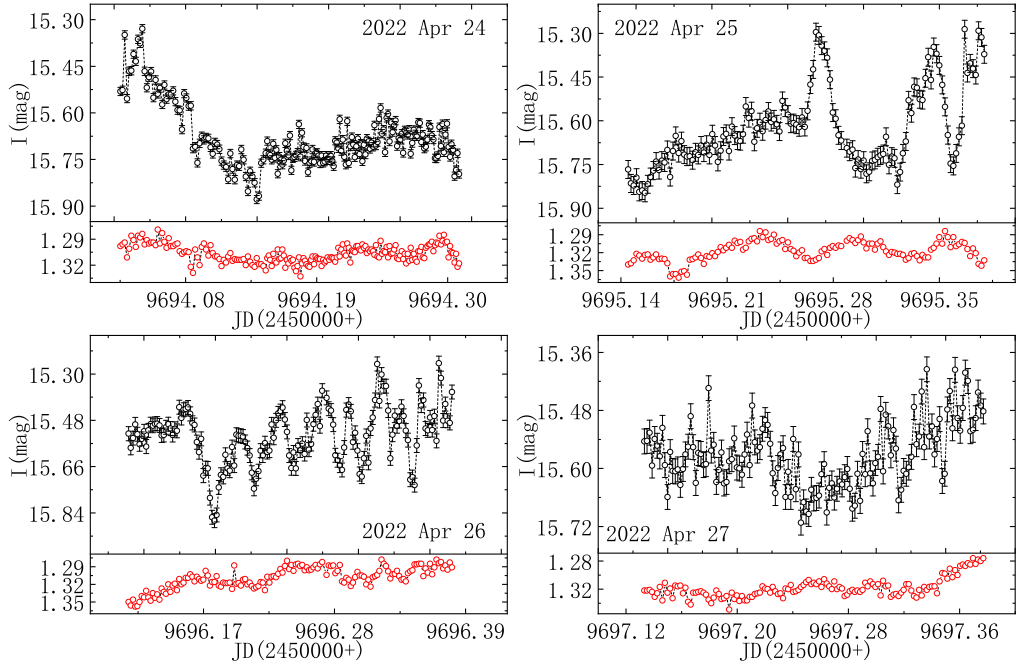

**Figure 2.** Light curves of 1ES 1426+42.8 in 2022.

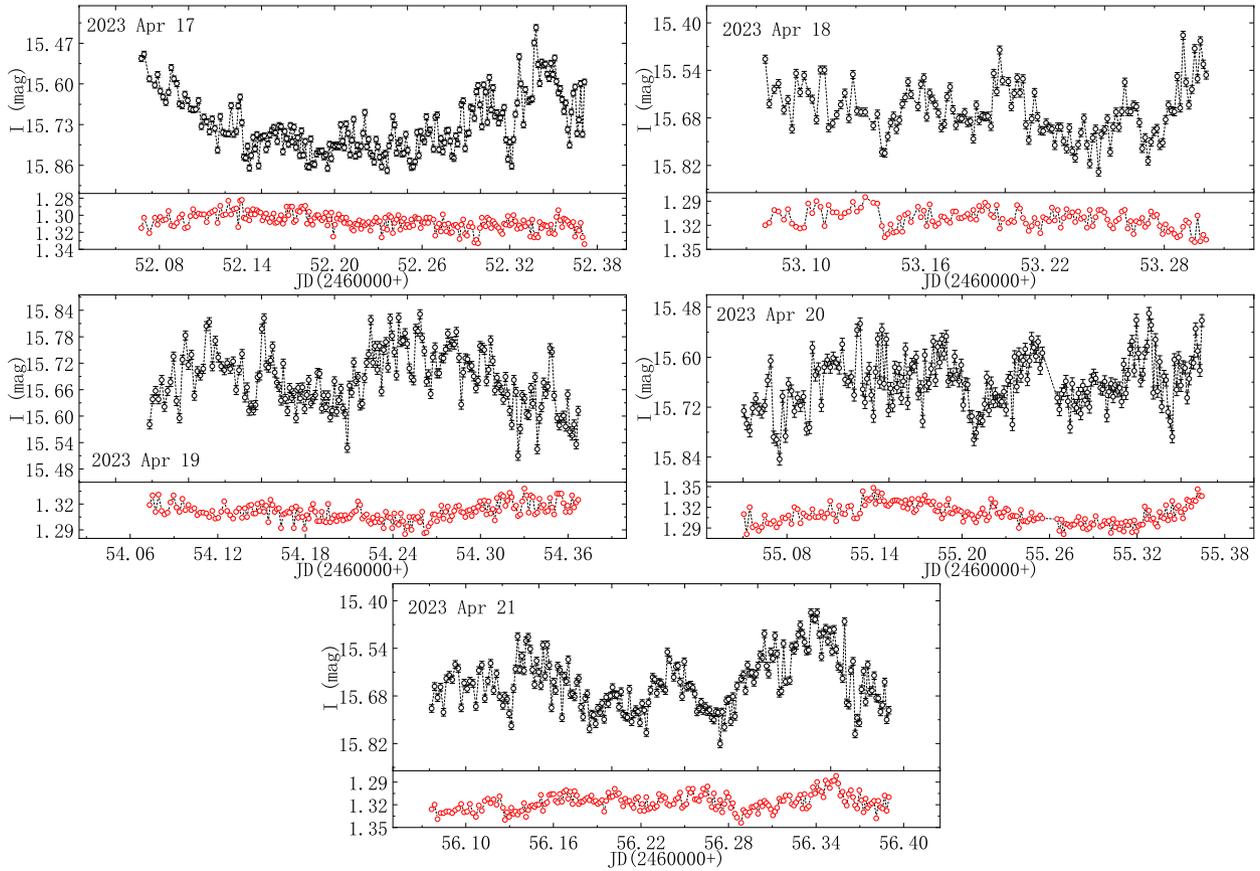

**Figure 3.** Light curves of 1ES 1426+42.8 in 2023.

**Table 2.** Results of IDV Observations. Column 1: the date of the observation; Column 2: the number of data points; Column 3: the $F$ value of Howell Test; Column 4: the critical $F$ value of Howell Test with 99 per cent confidence level; Column 5: the $F$ value of ANOVA; Column 6: the critical $F$ value of ANOVA with 99 per cent confidence level; Column 7: the variability status (V: variable, N: non-variable); Column 8: the daily average magnitudes and errors.

| Date | Number | $F$ | $F_c(99)$ | $F_a$ | $F_a(99)$ | V/N | Ave(mag) |
| --- | --- | --- | --- | --- | --- | --- | --- |
| 2021 Jan 25 | 51 | 0.08 | 1.95 | 1.39 | 2.58 | N | 15.576 (0.047) |
| 2021 Mar 16 | 64 | 3.81 | 1.81 | 3.31 | 2.32 | V | 15.594 (0.012) |
| 2021 Mar 23 | 13 | 1.86 | 4.16 | 15.23 | 7.01 | N | 15.867 (0.023) |
| 2021 Mar 24 | 42 | 2.68 | 2.09 | 9.95 | 2.84 | V | 15.490 (0.013) |
| 2021 Apr 22 | 42 | 0.20 | 2.09 | 1.62 | 2.84 | N | 15.326 (0.035) |
| 2021 Apr 29 | 20 | 0.04 | 3.03 | 2.60 | 5.29 | N | 15.581 (0.066) |
| 2021 May 18 | 42 | 0.02 | 2.09 | 1.40 | 2.84 | N | 16.129 (0.082) |
| 2022 Apr 24 | 195 | 2.76 | 1.40 | 5.44 | 1.63 | V | 15.681 (0.013) |
| 2022 Apr 25 | 127 | 2.32 | 1.52 | 16.84 | 1.82 | V | 15.627 (0.031) |
| 2022 Apr 26 | 146 | 1.56 | 1.47 | 6.31 | 1.75 | V | 15.539 (0.027) |
| 2022 Apr 27 | 133 | 0.81 | 1.50 | 6.27 | 1.80 | N | 15.573 (0.026) |
| 2023 Apr 17 | 219 | 2.51 | 1.37 | 7.44 | 1.58 | V | 15.731 (0.009) |
| 2023 Apr 18 | 137 | 1.27 | 1.49 | 5.43 | 1.78 | N | 15.661 (0.012) |
| 2023 Apr 19 | 189 | 1.25 | 1.41 | 21.41 | 1.64 | N | 15.683 (0.010) |
| 2023 Apr 20 | 193 | 0.64 | 1.40 | 20.40 | 1.63 | N | 15.660 (0.014) |
| 2023 Apr 21 | 205 | 1.67 | 1.39 | 5.20 | 1.61 | V | 15.642 (0.011) |

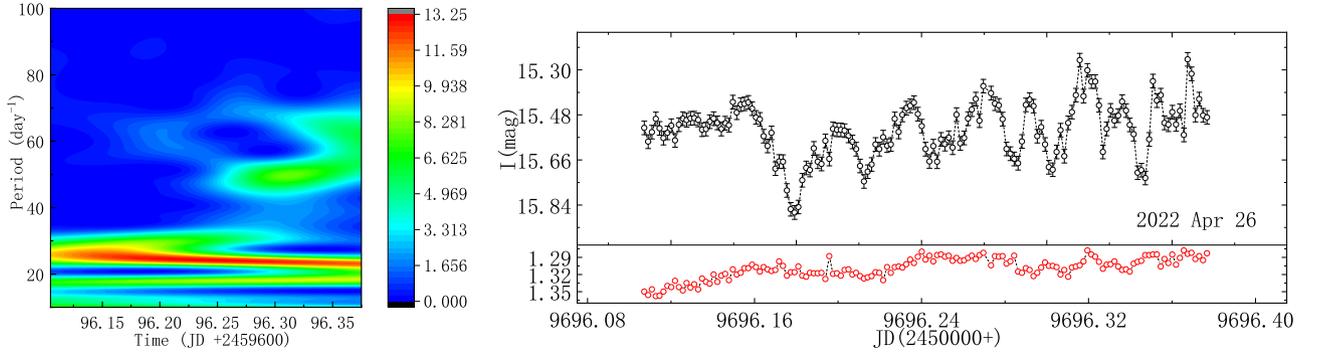

**Figure 4.** Left panel: the WWZ power for the whole light curve (right panel) on 2022 April 26. The bright red patch represents a possible QPO in the interval of JD(2459600+) 96.144 to 96.365 (segment 1). The brightest red patch represents a stronger signal between JD(2459600+) 96.144 and 96.286 (segment 2). Right panel: the light curve of 1ES 1426+42.8 on 2022 April 26.

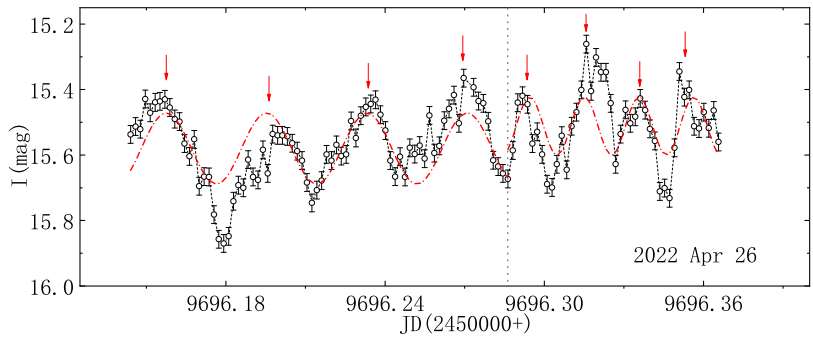

**Figure 5.** The light curve in the interval from JD(2459600+) 96.144 to 96.365 (segment 1). The interval before the grey vertical dashed line is JD(2459600+) 96.144 - 96.286 (segment 2), during which clear stable periodic modulation is seen. The red dash–dotted line represents the best-fitting curve of the sine function for this interval. The red arrows are plotted to indicate the modulation in this interval.

### 3.2.2 REDFIT method

The REDFIT program[3] (Schulz & Mudelsee 2002) based on the LSP is often used to estimate the QPOs and the red-noise levels in the light curves of blazars (Fan et al. 2014; Sandrinelli et al. 2016;

---
[3] http://www.geo.uni-bremen.de/geomod/staff/mschulz/

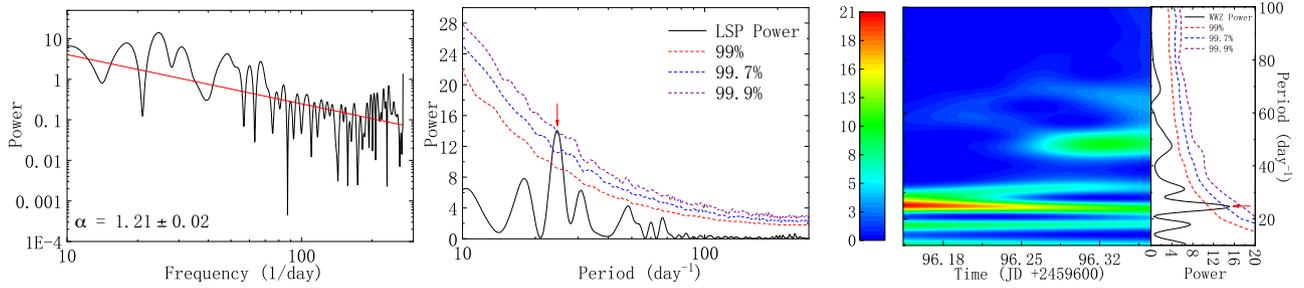

**Figure 6.** Period analysis results of segment 1. Left panel: the PSD of 1ES 1426+42.8 on 2022 April 26. The black solid line is the best-fitting power-law model of the underlying coloured noise. The red solid line is the best-fitting $P(f) \propto f^{-\alpha}$ profile to the black solid line ($\alpha = 1.21 \pm 0.02$). Middle panel: LSP of 1ES 1426+42.8. The black solid line represents the corresponding power spectrum of LSP. The red, blue, and purple dashed lines represent the confidence level of 99, 99.7 ($3\sigma$), and 99.9 per cent, respectively. The red arrow marks the period of the detected QPO. Right panel: 2D plane contour of the WWZ power. The black solid curve in the side panel represents the time-averaged WWZ power. The red, blue, and purple dashed lines represent the confidence level of 99, 99.7 ($3\sigma$), and 99.9 per cent, respectively.

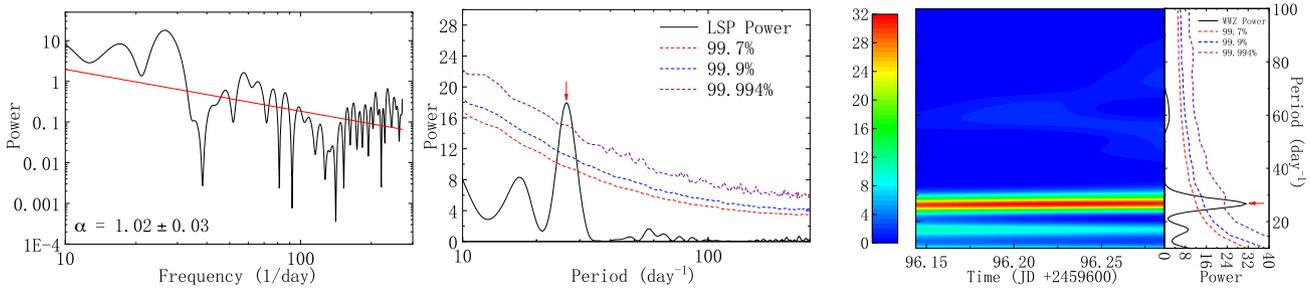

**Figure 7.** Similar plots of Fig. 7, but for segment 2.

Gupta et al. 2019). The REDFIT program can effectively remove the bias in the Fourier transform of unevenly spaced data by correcting for the correlation effect among Lomb–Scargle Fourier components. Additionally, the significances of peaks in the spectrum of unevenly spaced time series can be assessed against the red-noise background by fitting a first-order autoregressive (AR1) process (Xiong et al. 2017). The results of REDFIT are presented in the Fig. 8, where the bias-corrected power spectrum is shown by the black line, and the theoretical red-noise spectrum is shown by the red line. The significance levels of 90 per cent, 95 per cent, and 99 per cent based on Monte Carlo simulations are presented by the blue, green, and purple dashed lines, respectively. From Fig. 8, it can be seen that the peaks ($58.54 \pm 8.39$ min for segment 1 and $55.42 \pm 10.47$ min for segment 2, highlighted by the red arrow) in the spectrum of a time series are significantly detected ($> 99$ per cent) against the red-noise background.

In addition to the first four peaks in the segment 2 (QPO with a period of ∼55 min), we further conduct QPO analysis on the remaining four peaks ranging from JD(2459600+) 96.286 to 96.365 (Fig. 9). The result from the LSP method shows that there is a possible QPO with a period of $29.71 \pm 6.31$ min ($3\sigma$). Therefore, the QPO period is shortened from about 55 to 29.71 min. However, the remaining four peaks have lower significance and the shape of each cycle is more unstable compared to the first four peaks in segment 2.

## 4 DISCUSSION

We use different statistical methods to analyse the IDVs in the light curves of 1ES 1426+42.8. Among these IDVs, we found a potential QPO in the light curve on 2022 April 26. In the following subsections, we discuss the sub-hour IDV and possible QPO in terms of the potential explanations.

### 4.1 Variability theoretical model of IDV

The observed IDV/microvariability can be interpreted as a result of a convolution of individual synchrotron pulses occurring in the turbulent jets (Bhatta et al. 2013; Webb et al. 2021; Webb & Sanz 2023). In the relativistic jet model, if a shock interacts with a turbulent cell, it causes electrons in the cell to accelerate. The electrons then cool by synchrotron emission, ultimately resulting in the flare production (Bhatta et al. 2013; Webb & Sanz 2023). The synchrotron emission depends on the orientation and strength of the magnetic field, as well as the density of electrons (Webb 2016; Webb et al. 2021). The equation of Kirk et al. (1998; hereafter KRM) calculates the particle distribution in the shock front for various orientations of the magnetic field and particle densities (Kirk et al. 1998; Webb, Bhatta & Hollingsworth 2010). The burst profiles as described by the KRM model matched the pulse profiles of IDV/micro-variability observed in the actual blazar light curves (Bhatta et al. 2011). Using the model originally developed by Bhatta et al. (2013) and later improved by Xu et al. (2023), we recalculated the KRM burst profiles and compared them to our IDV/micro-variability light curves. The distribution of the particles is given by the diffusion equation:

$$\frac{\partial N}{\partial t} + \frac{\partial}{\partial \gamma}\left[\left(\frac{\gamma}{t_{acc}} - \beta_s \lambda^2\right) N\right] + \frac{N}{t_{esc}} = Q\delta(\gamma - \gamma_0), \quad (3)$$

where

$$\beta_s = \frac{4}{3}\frac{\sigma_t}{m_e c}\left(\frac{B^2}{2\mu_0}\right), \quad (4)$$

*Optical IDV analysis for 1ES 1426+42.8*    127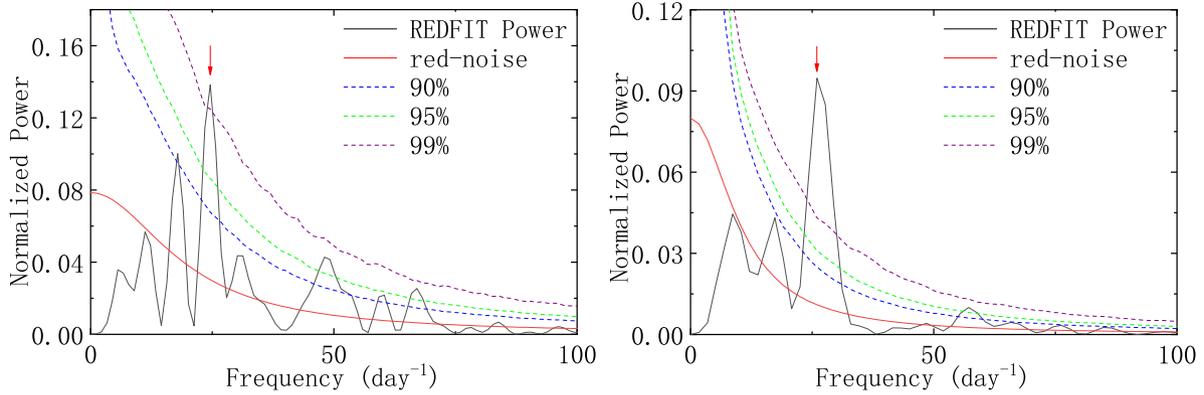

**Figure 8.** The REDFIT analysis on segment 1 (left) and segment 2 (right). The black solid line is the bias-corrected power spectrum. The red solid line is the theoretical red-noise spectrum. The blue, green, and purple dashed lines represent the confidence levels of 90, 95, and 99 per cent, respectively. The red arrow marks the period of the detected QPO.

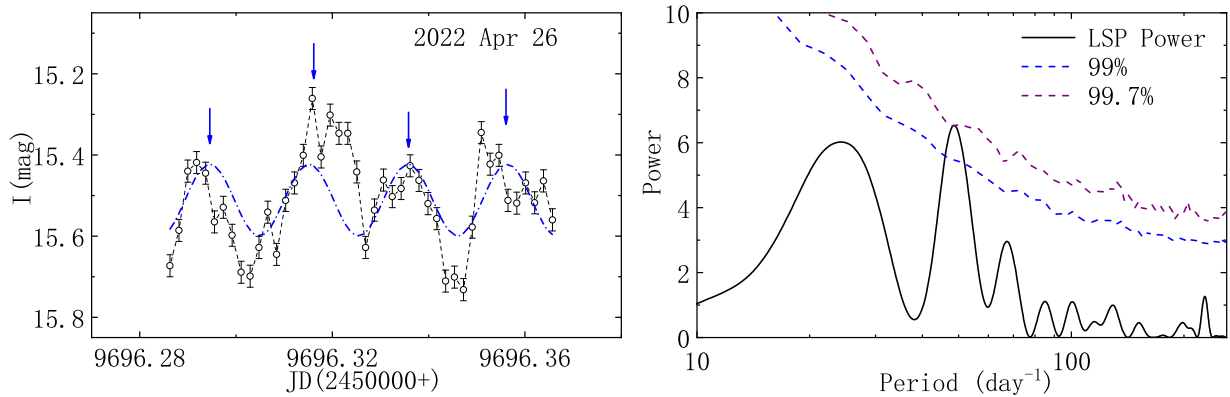

**Figure 9.** Left panel: the light curve of 1ES 1426+42.8 on 2022 April 26 from the interval JD(2459600+) 96.286 to 96.365. The blue dash–dotted line represents the best fit of the sine function for this interval. The blue arrows are plotted to indicate the modulation in this interval. Right panel: the LSP analysis of this interval. The black solid line represents the corresponding power spectrum of LSP. The blue and purple dashed lines represent the confidence level of 99 and 99.7 ($3\sigma$) per cent, respectively.

where $N$ is the number density of the electrons as a function of time t and energy $\gamma$. $t_{acc}$ and $t_{esc}$ are the particle acceleration time and the particle escape time, respectively. The ratio between them controls the pulse shape. $\beta_s$ is the synchrotron emission, where $B$, $\sigma_t$, $\mu_0$, $m_e$, and $c$ are the magnetic field strength, Thomsons scattering cross-section, permeability of the free space, mass of an electron, and the speed of light, respectively. The amplitude is determined by the parameter $Q$ and is related to the magnetic field strength $B$ and orientation $\theta$, as well as the enhanced electron density (Bhatta et al. 2013; Webb et al. 2021). KRM solves this equation for the case of a constant injection rate $Q_0$ after switch-on at $t = 0$.

The KRM solution determines the amplitude and shape of an individual pulse, which represents the emission from each individual turbulent cell (Webb et al. 2021). By adjusting the width and amplitude of the standard pulse to fit each significant pulse in the IDV light curves, the size of the emission region can be obtained (Webb et al. 2021; Webb & Sanz 2023). Fig. 10 shows the fitting results for the model. In each panel, the blue circles are observation data, the violet dashed lines indicate different flares obtained by the fit, and the solid black line is the total fit. The correlation coefficients of the light curves obtained by fitting the model were 0.81, 0.94, 0.92, 0.97, 0.91, 0.89, and 0.87, respectively. The resulting parameters for the pulses used to model the light curves are listed in Table 3. The first column shows the date of each IDV observation. Columns 2 and 3 give the number of pulses and the average amplitude. Column 4 shows the average width of the pulses ($\tau_{flare}$). Column 5 indicates the range of cell sizes in au based on the assumed shock speed of $0.1c$ and the duration of the pulse. The last column lists the correlation coefficients. The results indicate that turbulence exists in most regions of the jet. All turbulent cell sizes are within the range of 2.58 to 54.52 au, and the distribution of cell sizes is continuous. The smallest cell size typically corresponds to the Kolmogorov scale length of the turbulent plasma. Most of the dissipation occurs in the non-relativistic plasma, where the turbulent kinetic energy is dissipated into heat at the Kolmogorov scale. The largest cell size corresponds to either the size of the plasma jet or the correlation length within the plasma. Turbulent cells exceeding this scale may become unstable (Bhatta et al. 2013; Meng et al. 2017; Webb et al. 2021; Webb & Sanz 2023).

### 4.2 Magnetic reconnection model of QPO

The magnetic reconnection model has been proposed to explain the periodic variability of blazars (Dong et al. 2020). Dong et al. (2020) found QPO signatures in both the light curve and the polarization degree variations during their relativistic magnetohydrodynamic

Downloaded from https://academic.oup.com/mnras/article/533/1/120/7729077 by guest on 27 August 2024MNRAS **533**, 120–130 (2024)

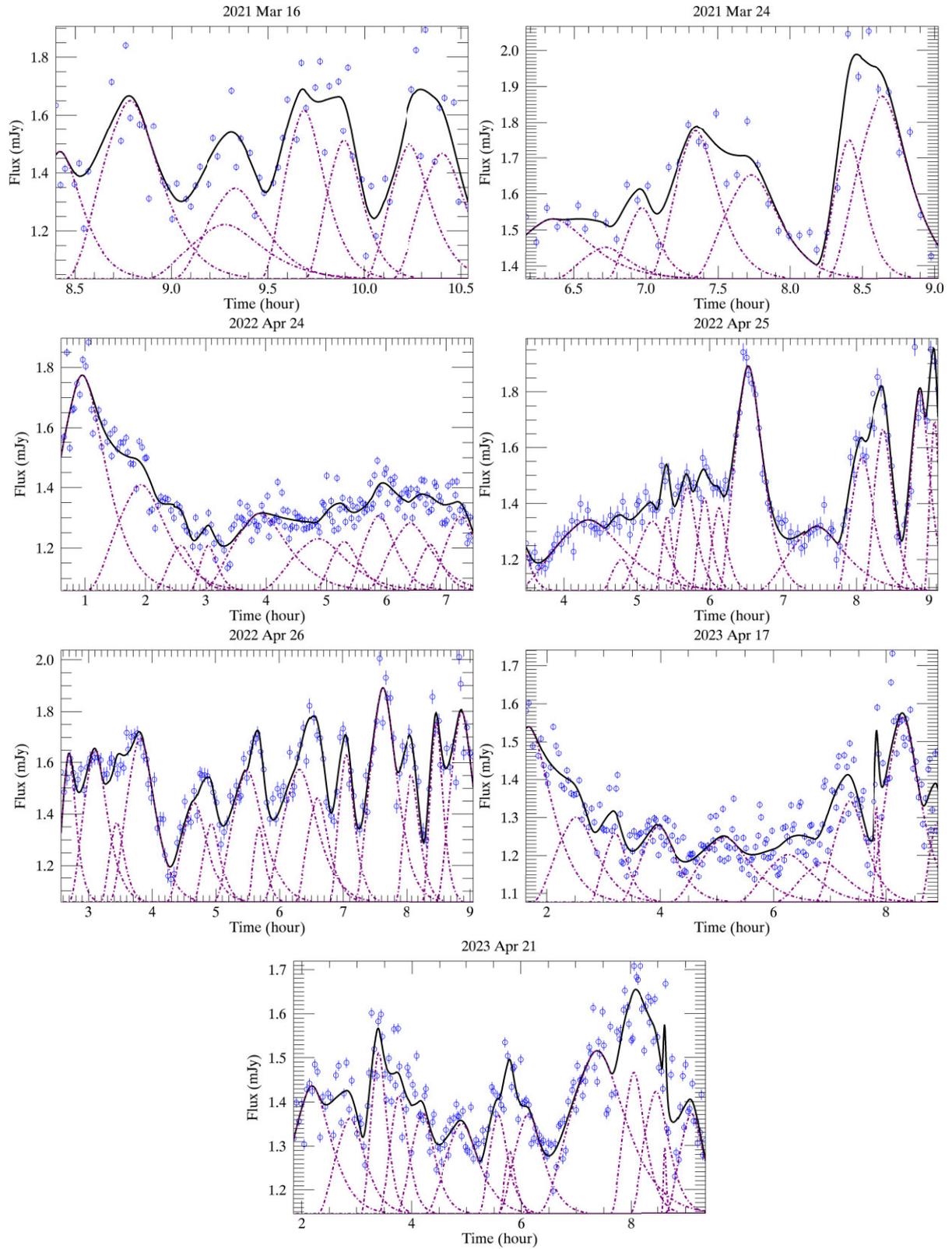

**Figure 10.** IDV fitting results. In each panel, the blue cycle points are the original data, the violet dashed lines show the fitted individual flares, and the black solid line is the fitting result.

**Table 3.** Pulse fit parameters.

| Date | Pulse number | Amplitude ($mJy$) | $\tau_{flare}$ (h) | Cell size (au) | Correlation coefficient |
| --- | --- | --- | --- | --- | --- |
| 2021 Mar 16 | 8 | 0.44 | 0.78 | 9.86–20.24 | 0.81 |
| 2021 Mar 24 | 7 | 0.29 | 1.01 | 11.40–26.98 | 0.94 |
| 2022 Apr 24 | 11 | 0.26 | 1.87 | 15.04–49.34 | 0.92 |
| 2023 Apr 25 | 14 | 0.39 | 1.02 | 6.99–50.27 | 0.97 |
| 2024 Apr 26 | 15 | 0.52 | 1.02 | 9.81–29.53 | 0.91 |
| 2023 Apr 17 | 11 | 0.24 | 1.83 | 4.16–48.29 | 0.89 |
| 2023 Apr 21 | 14 | 0.25 | 1.38 | 2.58–54.52 | 0.87 |

simulations. Recent observations from the Whole Earth Blazar Telescope also displayed intense polarimetry variability (Webb et al. 2021). These findings indicate significant periodic energy releases caused by kink instabilities occurring in a small subregion of the relativistic jets. The current-driven kink instability will be triggered when the jet is perturbed by a lateral displacement (Dong et al. 2020; Jorstad et al. 2022). The distorted magnetic field lines in the kinks disrupt the magnetic field structure, causing magnetic reconnection, accelerating particles, and further disturbing the magnetic field (Dong et al. 2020; Acharya, Borse & Vaidya 2021). The increase in poloidal field components and the number of accelerated particles maintain the polarization oscillation and outburst of non-thermal radiation caused by the kink instability (Zhang et al. 2017; Bodo, Tavecchio & Sironi 2021; Jorstad et al. 2022). The kink in the jet naturally evolves into a quasi-periodic structure, consisting of twisted magnetic fields (kink nodes). The process leads to the moving region (plasma blob) of enhanced emission, which contains a few kink nodes (Barniol Duran, Tchekhovskoy & Giannios 2017; Dong et al. 2020). Due to the quasi-periodic nature of the kink, this blob can generate rapid and distinct QPO radiation (Acharya et al. 2021; Jorstad et al. 2022; Raiteri et al. 2023). The period is associated with the kink growth time. The kink growth time can be estimated by the evolution of the kink's transverse motion (Mizuno et al. 2009). The kink's transverse displacement is roughly equal to the size of the emission blob (Jorstad et al. 2022). Then the period can be estimated as the ratio of the transverse displacement of the jet from its central spine over the averaged transverse velocity. Based on this, the period of QPOs expected from a kink can be estimated as follows: $P_{obs} = R_{KI}/v_{tr}\delta$, where $R_{KI}$ is the size of the emission region in the comoving frame of the jet (transverse displacement of the strongest kinked region), and $v_{tr}$ is the average transverse velocity of the kink (Dong et al. 2020). Using the typical transverse velocity $v_{tr} \sim 0.16c$ (Dong et al. 2020), the Doppler factor of the jet $\delta = 27.3$ (Wolter et al. 2008), along with our estimated upper limit of the emission region $8.15 \times 10^{12}$ $m$ (from the autocorrelation function, ACF, analysis, see Alexander 1997 for more details of the ACF), the period should then be $\lesssim 103.66$ min in the observer's frame. The possible QPO agrees well with this period upper limit. It should be noted that kink instabilities are not persistent physical processes in relativistic jets. The energy injected into the kinked jets can vary over time resulting in kink instabilities (Dong et al. 2020). The unchanged *I*-band flux intensity before the periodic behaviour of the light curve ($JD < 2459696.144$, see the right of Fig. 4) could be the period before the formation of kink. Once kink produced in the jet, the periodic modulation becomes apparent (the first four peaks in the segment 2). As the propagated kink in the jet partially dissipates or changes, the size of the emission region $R_{KI}$ becomes smaller, and then the period appears to be shortening phenomenon and the quasi-periodic variability of flux could become less obvious (the remaining four peaks after $JD = 2459696.286$). To sum up, the magnetic reconnection model in jets provides a good explanation for the possible short-lived QPO and the period shortening phenomenon.

## 5 CONCLUSION

Our main results are summarized as follows:

(1) We observed the BL Lac object 1ES 1426+42.8 for 16 nights from 2021 to 2023. The IDV is detected in 7 d.

(2) We applied the turbulent cell model to the optical observations of blazar 1ES 1426+42.8. We fit the light curves using this model and determine the distribution of cell sizes, amplitudes, and the width of pulses.

(3) The analysis yielded 80 pulses. The size distributions of the radiation turbulent cells, calculated from the fitting results, range from 2.58 to 54.52 au, which show that the distribution of cell sizes is consistent with the Kolmogorov distribution.

(4) A possible QPO with a period shortening phenomenon is found on 2022 April 26, which can be best explained by the magnetic reconnection model in the jet.

## ACKNOWLEDGEMENTS

This work is supported by the National Natural Science Foundation of China (grants 11863007, 12063005, 12063007,11703078), the Yunnan Province Foundation (2019FB004), the Program for Innovative Research Team (in Science and Technology) in University of Yunnan Province (IRTSTYN), and Yunnan Local Colleges Applied Basic Research Projects (2019FH001-12). We acknowledge the science research grants from the China Manned Space Project with NO. CMS-CSST-2021-A06. XC, CXL, and XWL acknowledge supports from the 'Science & Technology Champion Project' (202005AB160002) and from two 'Team Projects' – the 'Innovation Team' (202105AE160021) and the 'Top Team' (202305AT350002), all funded by the 'Yunnan Revitalization Talent Support Program'. This work is partially supported by a program of the Polish Ministry of Science under the title 'Regional Excellence Initiative', project no. RID/SP/0050/2024/1.

## DATA AVAILABILITY

The data underlying this article will be shared on reasonable request to the corresponding author.

## SUPPORTING INFORMATION

Supplementary data are available at *MNRAS* online.

**suppl_data**

Please note: Oxford University Press is not responsible for the content or functionality of any supporting materials supplied by the authors. Any queries (other than missing material) should be directed to the corresponding author for the article.

This paper has been typeset from a TEX/LATEX file prepared by the author.